\documentclass{article}

\usepackage[final]{neurips_2026}
\makeatletter
\renewcommand{\@noticestring}{Workshop on TAE (Trust-AI-Eval): Can We Trust AI Evaluation?, at 40th Conference on Neural Information Processing Systems (NeurIPS 2026).}
\makeatother
\workshoptitle{TAE (Trust-AI-Eval): Can We Trust AI Evaluation?}

\usepackage[utf8]{inputenc}
\usepackage[T1]{fontenc}
\usepackage{hyperref}
\usepackage{url}
\usepackage{booktabs}
\usepackage{amsfonts}
\usepackage{nicefrac}
\usepackage{microtype}
\usepackage{graphicx}
\usepackage{float}
\usepackage{tabularx}
\usepackage{array}
\usepackage{amsmath}
\usepackage{listings}
\usepackage[most]{tcolorbox}
\usepackage{longtable}

\graphicspath{{media/}}

\title{LLM-Generated Investment Research Memos Need Validation}

\author{%
  Chenri Luo$^{1}$\thanks{Equal contribution.}, \quad Xukang Gong$^{2,*}$, \quad Binyan Yu$^{1}$, \quad Xiao-Yang Liu$^{1}$\thanks{Corresponding author.}\\[4pt]
  \normalfont
  $^{1}$Columbia University,
  $^{2}$New York University\\
  Email: \{cl4575, by2445, xl2427\}@columbia.edu, xg2489@nyu.edu
}

\begin{document}

\maketitle

\begin{abstract}
LLMs are increasingly adopted to draft long-form investment research memos, but their reliability is judged claim by claim, while readers read each memo as a whole. In this paper, we show that a low claim-level error rate does not necessarily mean that a memo is reliable: even with a refutation rate of just 1–2\%, roughly four in ten memos contain at least one claim contradicted by official filings. First, we generate 1,440 memos with four recent LLM-powered agents for 30 companies over 12 quarters, decompose them into 77,272 financial claims, and verify each against the same official source bundle by retrieving evidence, matching it to the claim, and verifying the outcome. Second, we find that only 1.17--1.77\% of claims are refuted, yet 36.94--41.11\% of memos contain at least one refuted claim, and agents differ in severity: Codex has the lowest memo-level rate; a qualitative review further groups errors into four patterns (time, unit, context, concept).
\end{abstract}

\section{Introduction}

As LLMs have become more capable, their use in investment research has moved from experimentation to routine deployment: financial agents now take company filings as input and produce analyst-style memos end to end, with no human drafting step in between \citep{anthropic2026financial,xiao2026financeharness}. A system that is right most of the time still forces the reader to check every fact, because it does not say which ones are wrong \citep{churchman2026building}; professionals in practice review AI output selectively, focusing on the parts where it signals low confidence \citep{choi2026humanai}. A generated memo offers no such signal. In a Codex-generated Home Depot FY2024 Q4 memo (Appendix~\ref{app:memo_example}), comparable sales growth is reported as ``not disclosed'' in three sections, although the earnings release reports a 0.8\% increase.

Long-form factuality is typically evaluated at the claim level \citep{min2023factscore,wei2024longform,song2024veriscore}, and financial evaluation does the same, scoring individual questions \citep{islam2023financebench} or numerical claims \citep{hall2026verifin}. A claim-level rate answers ``how often is a randomly chosen statement wrong?'', whereas a reader asks ``does this memo contain a statement I should not rely on?'' With dozens of claims per memo, a low refutation rate says little about the latter. We call a memo \emph{contaminated} if it contains at least one refuted claim, i.e., one directly contradicted by aligned evidence in the official disclosures. Contamination is not about how serious the error is. It simply flags a memo that the reader should double-check, since there is no way to tell which claim is wrong just by reading it.

In this paper, we provide an empirical evaluation of LLM-generated investment memos and our key finding is that these memos need validation. Four LLM agents generate 1,440 memos for 30 public companies over 12 quarters. We decompose them into 77,272 financial claims and verify each against the same official source bundle used for generation, after aligning company, period, metric definition, and accounting basis, as shown in Fig.~\ref{fig:workflow}. Labels therefore show whether a claim matches the selected disclosures, not whether it is true in some absolute sense.

We address two research questions. \textbf{RQ1 (exposure):} does a low claim-level refutation rate imply few contaminated memos? No: only 1.17--1.77\% of claims are refuted, yet 36.94--41.11\% of memos are contaminated, and agents differ in how many refuted claims a contaminated memo contains (Section~\ref{sec:exposure}). \textbf{RQ2 (error patterns):} what do refuted claims look like, and what do the labels miss? They fall into time, unit, context, and concept errors; claims about post-period events may be labeled unverified rather than refuted, so refutation counts can understate risk (Section~\ref{sec:qualitative_patterns}).

Our main contributions are the development of three complementary reliability metrics and a \textbf{source-bounded validation pipeline}. The metrics include claim-level refutation, memo-level contamination, and error severity. They capture different aspects of reliability and should be reported together. The pipeline supports memo generation, claim extraction, and verification against official disclosures, with human audits used to assess the extraction and verification stages.

\section{Related Work}

\paragraph{Financial LLMs and benchmarks.}
Domain-adapted models, from FinBERT to BloombergGPT and FinGPT \citep{araci2019finbert,wu2023bloomberggpt,liu2023fingpt,liu2024fingpthpc}, have established finance as a major application area for LLMs. Hallucination has also been recognized as a key limitation of these models \citep{kang2023hallucination}. Existing financial benchmarks primarily evaluate short-form responses in tasks such as question answering, information retrieval, and reasoning \citep{islam2023financebench,xie2024finben,peng2025multifinben,hu2025finsearchcomp}. The closest to our setting is FinanceGym, introduced alongside the FinanceHarness framework \citep{xiao2026financeharness}, which evaluates long-form financial research reports against expert-defined rubrics. Even frontier systems achieve less than half of the available credit. However, rubric-based evaluation primarily measures the quality and completeness of a report's analysis, rather than whether its claims contradict the sources provided for generation. We focus on this complementary dimension of reliability.

\paragraph{Long-form factuality.}
FActScore \citep{min2023factscore}, SAFE \citep{wei2024longform}, and VeriScore \citep{song2024veriscore} decompose long-form output into claims and check their support against a knowledge source or search results. They separate supported from unsupported claims and summarize factual precision as an average, which does not show how many outputs contain at least one error. We verify against the official bundle supplied for generation, distinguish direct contradiction from unresolved evidence, and retain each claim's memo link, so the same labels yield both claim-level rates and memo-level exposure.

\paragraph{Financial verification.}
FinDVer~\citep{zhao2024findver} verifies claims against long financial documents with text and tables. FAITH~\citep{zhang2025faith} targets hallucination in tabular reasoning. FinDeepResearch~\citep{zhu2025findeepresearch} scores analytical quality with expert criteria. Hall et al.~\citep{hall2026verifin} verify and repair individual numerical answers with solver feedback. These works verify benchmark-provided claims, score analysis, or repair single answers; we instead extract claims from model-written memos and measure how source-aligned contradictions accumulate within each memo.

\section{Generation of Investment Research Memos}

\subsection{Company-period Cases and Source Boundary}
\label{sec:company_cases}

Our evaluation covers 30 public companies across 10 sectors, with three companies per sector, over 12 consecutive quarters from 2023 Q1 to 2025 Q4. This yields 360 company-quarter observations. The repeated quarterly design requires agents to handle changing financial results, reporting structures, and period-specific evidence. This prevents them from relying on a single static company profile. The company set spans ten sector categories---information technology, semiconductors and AI infrastructure, communication services, financials, health care, consumer discretionary, consumer staples, industrials, energy, and materials---with the complete roster listed in Table~\ref{tab:company_roster} (Appendix~\ref{app:company_roster}).

For every case, we construct a source bundle from official materials associated with the target reporting period. A typical bundle contains the Form 10-Q or 10-K, the earnings release, a README describing the reporting period and available files, and presentations, financial supplements, or transcripts when available. Generation and verification are constrained to this same bundle. We exclude outside knowledge, web search, market prices, analyst reports, and files from other companies or periods.

\subsection{Memo Generation}

We evaluate four memo-generation agents: Codex powered by GPT-5.5, Claude Code powered by Claude Sonnet 4.6, DeepSeek powered by DeepSeek-V4-Pro, and Qwen powered by qwen3.8-flash with thinking mode. Codex and Claude were accessed in June 2026 through agent-style command-line interfaces; DeepSeek and Qwen were both accessed in August 2026, through scripted single-call API generation with the source bundle placed directly in the prompt. We use \emph{agent} for these four configurations without implying identical interaction or reasoning procedures. The prompt follows an analyst-style equity research structure informed by CFA Institute guidance~\cite{cfa2020equity}. It requires seven sections: Executive Summary / Investment View, Business Segment Performance, Management Guidance and Strategic Commentary, Key Metrics Table, Risks and Concerns, Forward-Looking Signals, and Source-Grounded Conclusion. It prohibits unsupported figures, guidance, valuation assumptions, ratings, and price targets; requires GAAP/non-GAAP distinctions; and permits causal language only when supported by the sources.

Each memo is constrained to 900--1,200 words, excluding a 6--8 item metrics table. All four agents operate under the same document-length and structural requirements. They nevertheless differ in claim density (claims per memo; Table~\ref{tab:evaluation_scale}). We treat this difference as part of agent behavior: a reader receives the complete memo produced under the common specification, not a version later adjusted to make claim counts equal across agents.

\begin{figure}[t]
  \centering
  \includegraphics[width=0.88\linewidth]{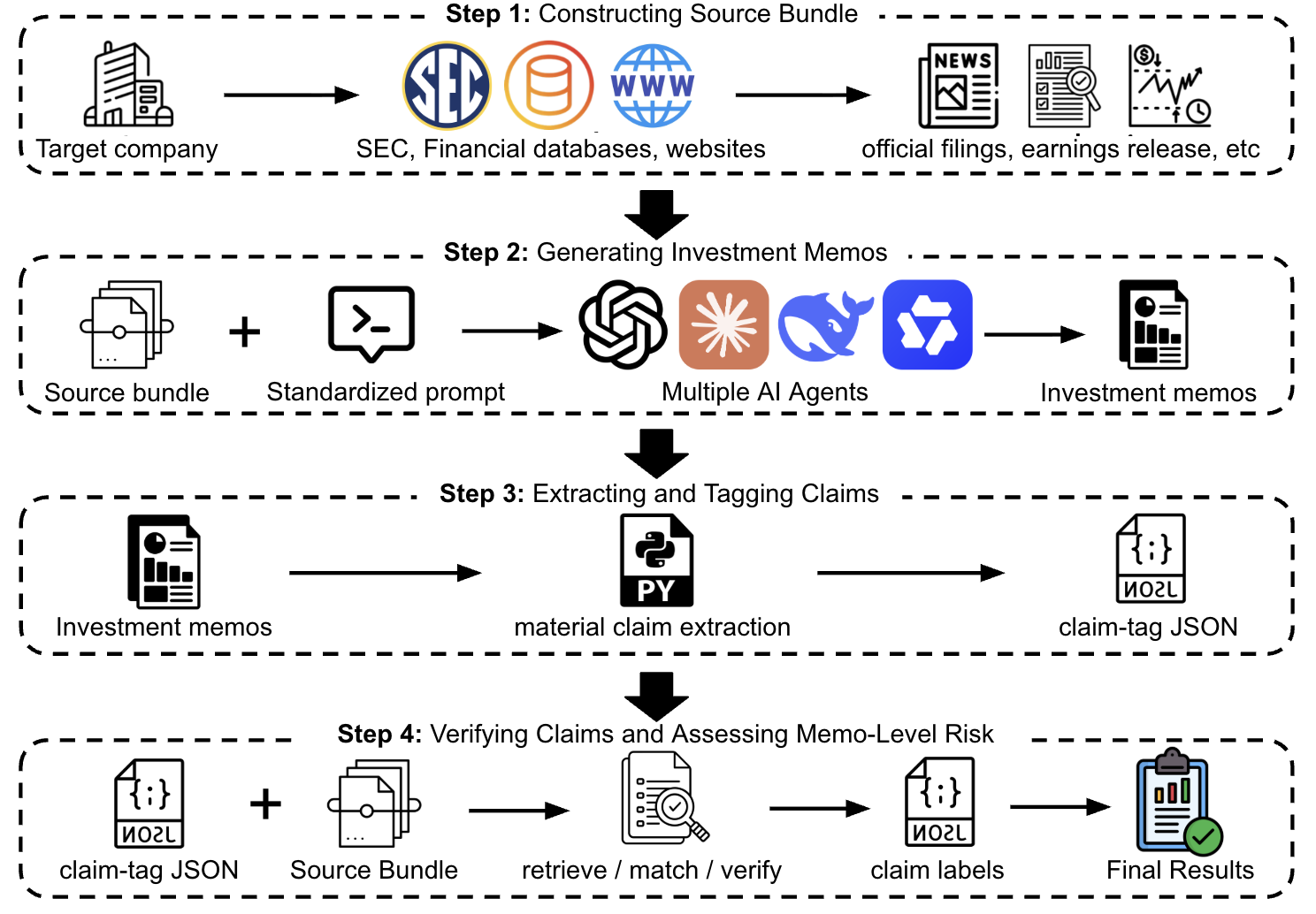}
  \caption{Overview of the source-bounded memo generation and validation workflow.}
  \label{fig:workflow}
\end{figure}

Figure~\ref{fig:workflow} summarizes the source-bounded generation and validation workflow. The same fixed company-period bundle is used for memo generation and claim verification, while the human audit evaluates the automated verifier rather than redefining the claim labels.

\begin{table}[H]
\centering
\small
\caption{Evaluation scale and extracted claims.}
\label{tab:evaluation_scale}
\begin{tabular}{lrrrrr}
\toprule
Agent & Memos & Claims & Claims/memo & Min & Max \\
\midrule
Codex & 360 & 22,757 & 63.2 & 45 & 80 \\
Claude & 360 & 17,920 & 49.8 & 22 & 85 \\
DeepSeek & 360 & 17,416 & 48.4 & 35 & 70 \\
Qwen & 360 & 19,179 & 53.3 & 41 & 73 \\
\bottomrule
\end{tabular}
\end{table}

\section{Evaluation Method}

The evaluation protocol is summarized as
\[
S \rightarrow M \rightarrow C(M) \rightarrow V,
\]
where $S$ is a company-period source bundle, $M$ is the generated memo, $C(M)=\{c_1,\ldots,c_t\}$ is its induced set of material claims, and $V$ is the set of source-based verification outcomes. The protocol separates three questions: what the memo claims, which evidence is aligned to each claim, and what conclusion that evidence supports.

\subsection{Claim Extraction and Routing}

The claim tagger extracts material statements from prose and metrics tables and emits one structured JSON record per claim. Records retain the memo excerpt, section and location, company-period metadata, semantic claim kind, verification route, and structured numeric fields when available. Material claims include current-period values, comparisons, computed changes, causal attributions, management guidance, descriptive facts, analytical judgments, and source-boundary statements.

Claim type determines the initial evidence operation. Current-period values use official fact-store matching; comparisons use aligned prior-period values; computed changes are recalculated; narrative and causal statements use source-text lookup; guidance uses official outlook materials; and table rows may use same-table lookup or recomputation. An example structured claim record appears in the appendix.

\subsection{Source-based Verification}
\label{sec:source_based_verification}

The automated verifier retrieves candidate evidence from tables and text in the company-period source bundle, including an internal fact store built from those documents. Before using a candidate to support or refute a claim, it checks alignment in company, reporting period, metric definition, accounting basis, reporting scope, units, and source version. These are rule-based checks on the extracted evidence, not a test of XBRL tag identity. Quarterly values are not treated as interchangeable with full-year or year-to-date values, nor are GAAP and non-GAAP measures (see Appendix~\ref{app:gaap-xbrl}). A mismatch in period, scope, metric, or accounting basis does not by itself establish a contradiction.

Each claim receives one of five mutually exclusive labels. \texttt{verified} means that aligned evidence supports the material content. \texttt{partially\_verified} means the central statement is supported but one or more values, qualifiers, comparisons, or details cannot be fully established. \texttt{not\_applicable} covers content outside direct factual verification. \texttt{unverified} means a factually checkable claim cannot be resolved from the available evidence. \texttt{refuted} requires aligned evidence that directly contradicts a material part of the claim. Not-applicable and unverified claims are not counted as errors.

\subsection{Claim- and Memo-level Measures}
\label{sec:claim-memo}
For $N$ extracted claims, the rate for outcome $k$ is $N_k/N$. Our main local error measure is the refuted-claim rate because it requires a direct contradiction rather than missing evidence. At the memo level, a memo is contaminated if it contains at least one refuted claim. We also report refuted claims per memo, refuted claims per contaminated memo, and the distribution of 0, 1, 2, 3--5, and 6 or more refuted claims.

The contamination metric addresses a different evaluative question from claim-level accuracy. Claim-level rate estimates the frequency of contradictions among extracted statements. Contamination estimates whether a reader of the complete memo encounters at least one contradiction without knowing which claim it is. It is therefore a validation-trigger metric, not a claim that all content in a contaminated memo is false.

Claim count affects this exposure by construction. If each of $n$ claims were independently refuted with constant probability $p$, the probability of at least one refutation would be
\[
P(\text{contaminated}\mid n,p)=1-(1-p)^n.
\]
We do not use this independence model to predict the observed rates because errors may cluster by memo, company, metric, or verification route. Instead, it clarifies why the claim-level rate and memo-level exposure answer different questions. This relates to a broader caution for AI-agent traces: even when each individual action is permissible on its own, the full trace may not be \citep{liu2026nocertificate}; we discuss this relationship, and how it differs from our setting, in Appendix~\ref{app:related-theory}. The four agents were constrained to the same memo length and structure; their different claim counts reflect differences in output density and are reported alongside both rate- and memo-based measures.

\subsection{Quality Control and Human Audit}
\label{sec:human_audit}

This subsection reports separate audits of claim extraction and automated verification. We manually audited 500 extracted claim records across all four agents, all 30 companies, and all major memo sections: claim boundaries were correct in 96\% of cases, 100\% preserved the intended memo meaning, and 91\% received the correct claim type, with most remaining errors involving borderline segmentation or closely related claim types.

We also audited the automated verifier (\texttt{verify\_claims.py}). For the largest outcome class, we sampled 500 records labeled \texttt{verified}; a numeric-presence screen flagged 22 for closer inspection, and we separately reviewed 120 records randomly sampled from the remaining 478. Among the 22 screen-flagged records, [X1] confirmed the automated decision, [Y1] contained a labeling error, and [Z1] were borderline. Among the 120 randomly sampled records, [X2] confirmed, [Y2] contained a labeling error, and [Z2] were borderline. We report the two groups separately because the screen-flagged records were not randomly sampled, so pooling them would overstate the error rate; the random sample alone gives an error-or-borderline rate of [(Y2+Z2)/120]\%.

Because contamination depends directly on the \texttt{refuted} label, we separately reviewed 200 refuted claims, 50 per agent. Human review confirmed 188 (94.0\%): 47 of 50 Codex, 46 of 50 Claude, 48 of 50 DeepSeek, and 47 of 50 Qwen. Most disagreements involved reporting-period alignment, metric definition, or evidence that was relevant but insufficient for direct contradiction.

These audits evaluate whether the pipeline applies its labels consistently within the defined source boundary; they do not establish objective financial truth outside the selected evidence set.

\section{Results}

\subsection{Claim-level Outcomes}

Table~\ref{tab:claim_verification_outcomes} reports all 77,272 outcomes. Codex has the highest verified rate and lowest unverified rate, whereas Claude has the highest partially verified rate. DeepSeek and Qwen fall between them on these measures; DeepSeek has the lowest refuted rate, followed closely by Qwen. Explicit refutation is rare: 402 of 22,757 Codex claims (1.77\%), 229 of 17,920 Claude claims (1.28\%), 204 of 17,416 DeepSeek claims (1.17\%), and 228 of 19,179 Qwen claims (1.19\%).

\begin{table}[H]
\centering
\small
\caption{Claim-level verification outcomes; cells report count and rate.}
\label{tab:claim_verification_outcomes}
\resizebox{\textwidth}{!}{%
\begin{tabular}{lrrrrrr}
\toprule
Agent & Claims & Verified & Partially verified & N/A & Unverified & Refuted \\
\midrule
Codex & 22,757 & 10,779 (47.37\%) & 6,045 (26.56\%) & 4,350 (19.11\%) & 1,181 (5.19\%) & 402 (1.77\%) \\
Claude & 17,920 & 6,679 (37.27\%) & 7,613 (42.48\%) & 1,505 (8.40\%) & 1,894 (10.57\%) & 229 (1.28\%) \\
DeepSeek & 17,416 & 7,974 (45.79\%) & 6,175 (35.46\%) & 1,684 (9.67\%) & 1,379 (7.92\%) & 204 (1.17\%) \\
Qwen & 19,179 & 8,446 (44.04\%) & 7,181 (37.44\%) & 2,037 (10.62\%) & 1,287 (6.71\%) & 228 (1.19\%) \\
\bottomrule
\end{tabular}}
\end{table}

The refuted claims also reveal different failure profiles. Incorrect numerical values account for 245 Codex refutations (60.95\%), 180 Claude refutations (78.60\%), 179 DeepSeek refutations (87.75\%), and 170 Qwen refutations (74.56\%). Incorrect growth calculations account for 25 (6.22\%), 49 (21.40\%), 25 (12.25\%), and 56 (24.56\%), respectively. Codex additionally produces 132 incorrect statements that a metric was not disclosed even though matching disclosure exists in the source bundle. Qwen produces two such statements, while neither Claude nor DeepSeek produces any.

\subsection{From Local Errors to Memo-level Exposure}
\label{sec:exposure}
Table~\ref{tab:memo_contamination} gives the central result. Although all four refuted-claim rates are below 2\%, contamination rates reach 36.94\% for Codex, 40.56\% for Claude, 40.28\% for DeepSeek, and 41.11\% for Qwen. Each of these rates means more than a third of memos contain at least one refuted claim. Codex therefore has a slightly lower observed contamination rate, but once a Codex memo is contaminated, it tends to be worse: 3.02 refuted claims on average, versus 1.57 for Claude, 1.41 for DeepSeek, and 1.54 for Qwen (Table~\ref{tab:memo_contamination}).

\begin{table}[H]
\centering
\small
\caption{Memo-level contamination and severity.}
\label{tab:memo_contamination}
\resizebox{\textwidth}{!}{%
\begin{tabular}{lrrrrrrr}
\toprule
Agent & Memos & Contaminated & Clean & Contam. rate & Refuted & Ref./memo & Ref./contam. \\
\midrule
Codex & 360 & 133 & 227 & 36.94\% & 402 & 1.12 & 3.02 \\
Claude & 360 & 146 & 214 & 40.56\% & 229 & 0.64 & 1.57 \\
DeepSeek & 360 & 145 & 215 & 40.28\% & 204 & 0.57 & 1.41 \\
Qwen & 360 & 148 & 212 & 41.11\% & 228 & 0.63 & 1.54 \\
\bottomrule
\end{tabular}}
\end{table}

\begin{figure}[t]
\centering
\includegraphics[width=\textwidth]{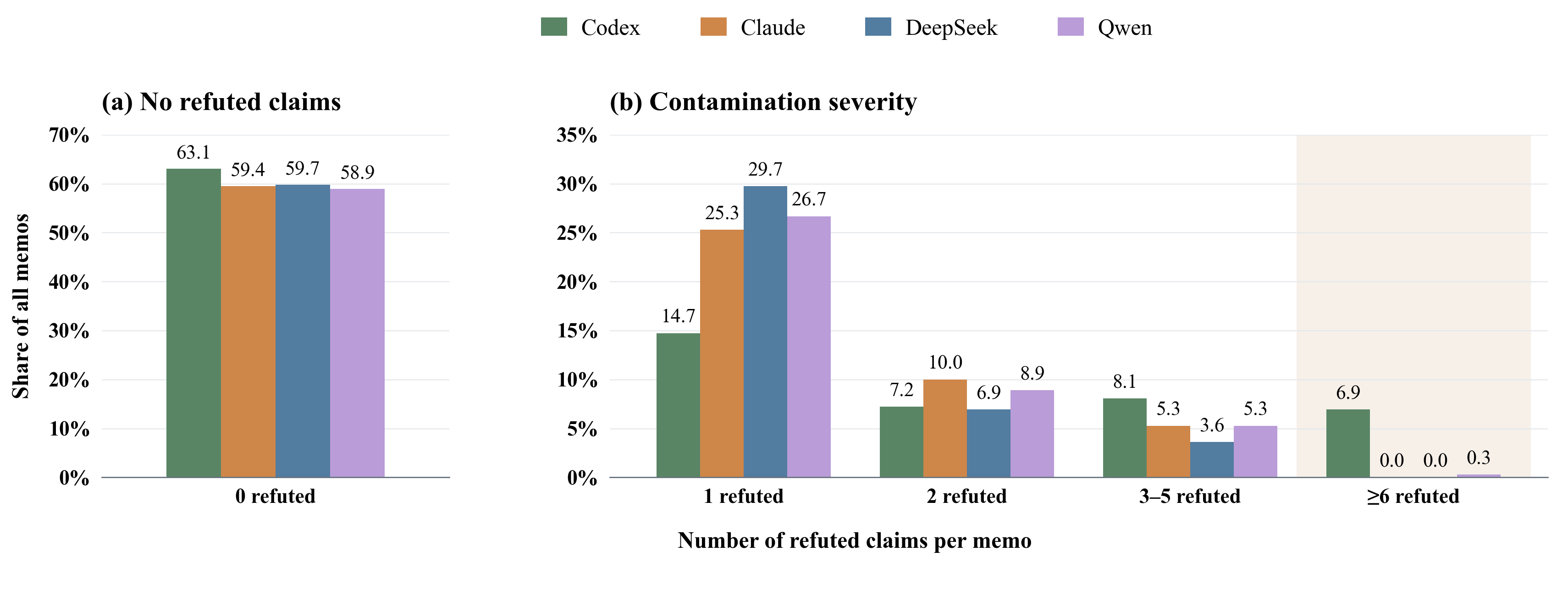}
\caption{Distribution of refuted claims across generated memos: (a) no refuted claims; (b) contamination severity. Percentages use all 360 memos per model; y-axis ranges differ between panels.}
\label{fig:contamination_severity}
\end{figure}

Figure~\ref{fig:contamination_severity} further separates exposure from severity. Codex produces more clean memos than the other three agents (63.1\% versus 58.9--59.7\%), while Claude, DeepSeek, and Qwen produce more memos with exactly one refuted claim. This pattern reverses at the high end: Codex has more memos with three or more refuted claims, including 25 memos (6.94\%) with six or more, compared with at most one such memo for any other agent. A binary contamination rate alone would miss this reversal. Claim density also differs systematically across agents; Appendix~\ref{app:claim_density} discusses how this interacts with the measures reported here.

\subsection{Verification Strategies and Evidence Operations}
We further examine not where refuted claims appear but how they are detected during source-grounded verification, with percentages computed within each agent's own refuted-claim set.

\begin{figure}[t]
\centering
\includegraphics[width=0.95\textwidth]{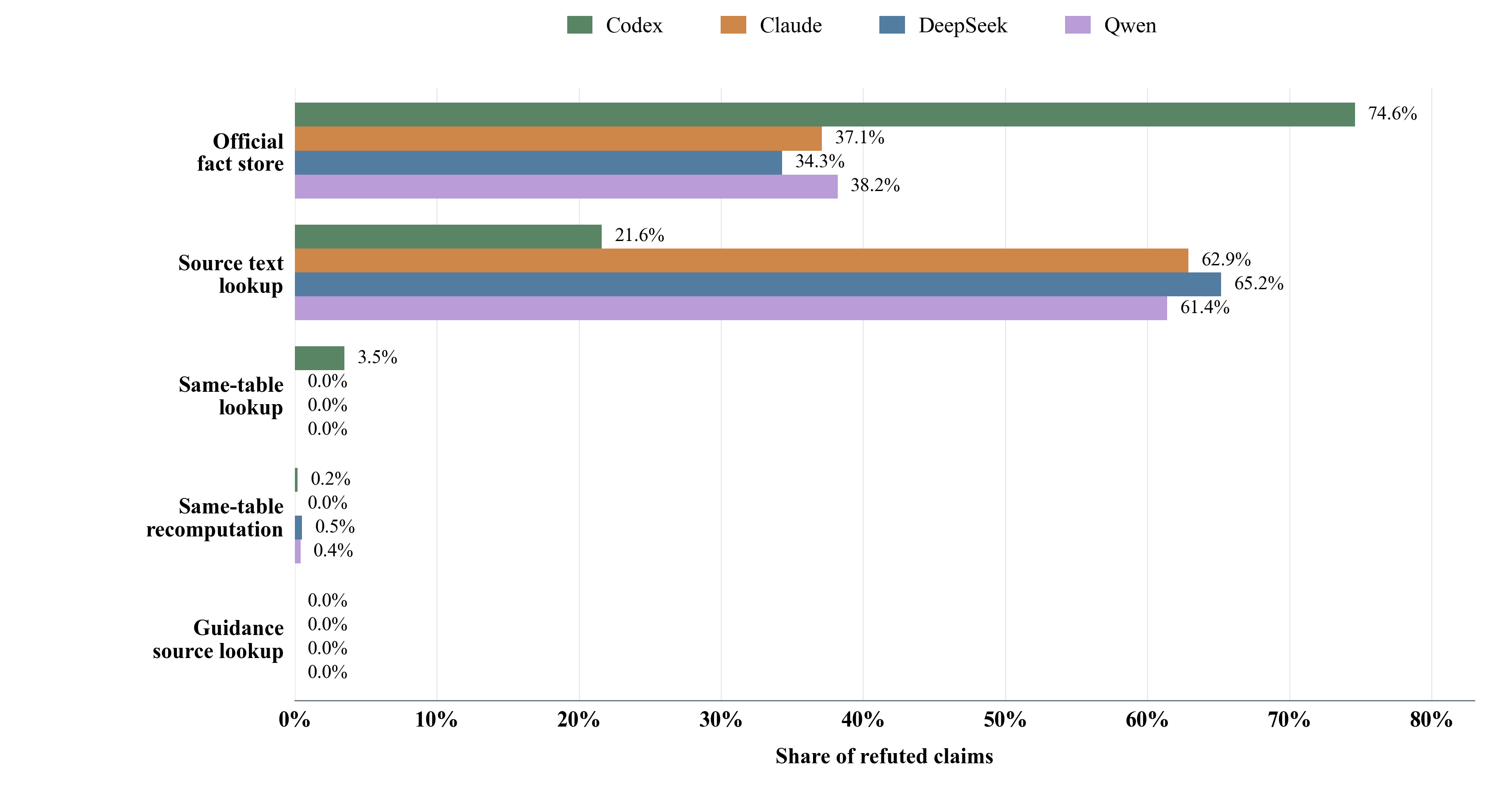}
\caption{Evidence operations used by the verifier to detect each agent's refuted claims.}
\label{fig:verification_strategies}
\end{figure}

Figure~\ref{fig:verification_strategies} reports the distribution. For Codex's refuted claims, the verifier relies primarily on official fact-store matching (74.6\%, versus 21.6\% for source-text lookup). For Claude, DeepSeek, and Qwen, the pattern reverses: source-text lookup dominates (62.9\%, 65.2\%, and 61.4\%, respectively, versus 37.1\%, 34.3\%, and 38.2\% for fact-store matching), reflecting a larger role for narrative disclosures. The remaining operations contribute little: guidance-source lookup identifies no refuted claims for any agent, same-table lookup appears only among Codex's refuted claims (3.5\%), and same-table recomputation identifies at most one claim per agent (0.2--0.5\%). These percentages show which verification operations were used most often for each agent. Since different operations identify different types of errors, the verifier combines them rather than relying on a single operation.

\subsection{Qualitative Error Patterns Beyond the Label Scheme}
\label{sec:qualitative_patterns}

The five verification labels defined in Section~\ref{sec:source_based_verification} record \emph{whether} a claim is supported, but not \emph{how} an unsupported claim came to be wrong. As a complementary qualitative check, separate from the stratified audits in Section~\ref{sec:human_audit}, we asked an investment practitioner to closely read individual generated memos against their source bundles. We identified four recurring patterns of errors. These patterns may overlap and are intended to illustrate common failure modes, not to provide a systematic re-annotation of the original corpus of 58,093 claims from three agents. The review did not include Qwen.

\textbf{Time.} A claim describes an event, status, or process that falls outside the source bundle's reporting period. This is particularly problematic when information about a later period is presented as an established fact. Such claims often risk being labeled \texttt{unverified} rather than \texttt{refuted}, since no single sentence in the bundle directly contradicts them, even when they anchor a memo's entire outlook.

\textbf{Unit.} A corresponding figure exists in the source, but its value or direction is stated incorrectly, as when a reported growth figure has the wrong sign or magnitude relative to the disclosed value.

\textbf{Context.} The model draws from the correct document but the wrong location within it, e.g., swapping two adjacent rows of a guidance table or substituting nearby descriptive language for the specific disclosed figure the prompt requires.

\textbf{Concept.} Two distinct financial concepts are treated as the same even though each is individually well defined and disclosed, such as using operating cash flow and free cash flow interchangeably.

These examples show that the verification labels alone do not fully explain how errors occur. Broader human review across verification labels is an important direction for future work, as discussed in the Limitations section.

\section{Limitations and Future Work}

The evaluation is source-bounded. A \texttt{verified} label means consistency with the selected official bundle, not objective or permanent truth. Official materials can differ in timing, accounting basis, reporting scope, and later amendment. Alignment rules reduce false comparisons but cannot ensure that the bundle contains all evidence needed to verify every claim.

Contamination is a conservative validation trigger, not a materiality score. One minor refutation differs from several central numerical errors; we therefore report severity distributions and error types. Agents also differ in where refuted claims are detected. Architecture and pretraining composition may contribute \cite{petty2025codepretrain}, but unavailable proprietary training details prevent direct testing.

Temporal leakage remains unresolved (Section~\ref{sec:qualitative_patterns}): reporting-period alignment does not detect every narrative reference to post-period events. Such claims may remain \texttt{unverified}, so refutation rates do not capture all temporal leakage. Rule-based evidence matching can also fail on ambiguous or variable phrasing, retrieving evidence from a mismatched period or a semantically unrelated sentence with similar keywords. This can mislabel an accurate claim. We observed such failures during manual tracing, including a quarterly figure checked against full-year evidence.

The study evaluates four complete agent configurations under one prompt in one domain. They differ in model family, version, access date, and invocation method, as detailed in Section~3.2; we do not isolate model-provider or country-of-origin effects. Generalization beyond the 30-company, 12-quarter design requires further evaluation. Extraction and verification introduce measurement error, requiring documentation and audits.

Future work has four directions. First, broader human review stratified by verification label and materiality-weighted severity measures could better connect refutations to reader risk. Second, the protocol and labeled dataset could support evaluations of runtime systems that trace claim provenance, check numerical accuracy, and assess institutional controls before presenting claims to readers.

Third, a drafting-time verify-and-revise loop could extend the current post-generation pipeline. Appendix~\ref{app:verify_revise_loop} outlines stopping conditions, revision budgets, and insufficient-evidence handling. This remains a research direction: we have not tested whether it reduces memo-level contamination.

Fourth, an LLM-based verifier could replace rule-based matching with case-by-case reasoning about period, scope, and metric alignment. This targets robustness to variable phrasing in the labeling process; the drafting-time loop instead targets revision of the memo itself.

\section{Conclusion}

We presented a source-bounded protocol for evaluating LLM-generated investment research memos at both claim and memo levels, with human audits of the automated pipeline. Across 1,440 memos and 77,272 claims, only 1.17--1.77\% of claims were refuted, yet 36.94--41.11\% of memos contained at least one refuted claim. Readers encounter complete memos and may rely on an incorrect statement without knowing which claim needs checking. Evaluations should therefore report claim-level errors, memo-level exposure, and severity together, while specifying the sources and reporting periods covered.

\section*{Acknowledgments}

Chenri Luo, Xukang Gong and Xiao-Yang Liu acknowledge support from Columbia University's SEAS Interdisciplinary Research (SIRS) and SEAS Translational Acceleration Research (STAR) Seed Funding Program and the Tang Family Fund for Research Innovations in FinTech, Engineering, and Business Operations, and the Center for Digital Finance and Technologies (CDFT). Xiao-Yang Liu acknowledges support from the JPMorganChase Faculty Research Award.

The opinions expressed in this paper are those of the authors and do not necessarily reflect those of JPMorganChase or its affiliates. This material is not a product of the Research Department of J.P. Morgan Securities LLC and does not constitute investment advice, a recommendation, or a solicitation or offer in any jurisdiction.

\clearpage
\bibliographystyle{plain}
\bibliography{references}

@inproceedings{kang2023hallucination,
  title={Deficiency of Large Language Models in Finance: An Empirical Examination of Hallucination},
  author={Kang, Haoqiang and Liu, Xiao-Yang},
  booktitle={Workshop on Failure Modes in the Age of Foundation Models (I Can't Believe It's Not Better) at NeurIPS 2023},
  year={2023}
}

@inproceedings{zhang2025faith,
  title={FAITH: A Framework for Assessing Intrinsic Tabular Hallucinations in Finance},   
  author={Zhang, Mengao and Fu, Jiayu and Warrier, Tanya and Wang, Yuwen and Tan, Tianhui and Huang, Ke-Wei},
  booktitle={Proceedings of the 6th ACM International Conference on AI in Finance},
  pages={159--167},
  year={2025},
  publisher={Association for Computing Machinery},
  address={New York, NY, USA},
  doi={10.1145/3768292.3770433}
}

@inproceedings{zhao2024findver,
  title={{FinDVer}: Explainable Claim Verification over Long and Hybrid-content Financial Documents},
  author={Zhao, Yilun and Long, Yitao and Jiang, Tintin and Wang, Chengye and Chen, Weiyuan and Liu, Hongjun and Tang, Xiangru and Zhang, Yiming and Zhao, Chen and Cohan, Arman},
  booktitle={Proceedings of the 2024 Conference on Empirical Methods in Natural Language Processing},
  pages={14739--14752},
  year={2024},
  publisher={Association for Computational Linguistics},
  doi={10.18653/v1/2024.emnlp-main.818}
}

@article{zhu2025findeepresearch,
  title={{FinDeepResearch}: Evaluating Deep Research Agents in Rigorous Financial Analysis},
  author={Zhu, Fengbin and Ng, Xiang Yao and Liu, Ziyang and Liu, Chang and Zeng, Xianwei and Wang, Chao and Tan, Tianhui and Yao, Xuan and Shao, Pengyang and Xu, Min and Wang, Zixuan and Wang, Jing and Lin, Xin and Li, Junfeng and Zhu, Jingxian and Zhang, Yang and Wang, Wenjie and Feng, Fuli and Hong, Richang and Luan, Huanbo and Huang, Ke-Wei and Chua, Tat-Seng},
  journal={arXiv preprint arXiv:2510.13936},
  year={2025}
}

@misc{cfa2020equity,
  author={{CFA Institute}},
  title={Equity Research Report Essentials},
  year={2020},
  howpublished={\url{https://www.cfainstitute.org/sites/default/files/-/media/documents/support/research-challenge/challenge/rc-equity-research-report-essentials.pdf}},
  note={Accessed: 2026-07-01}
}

@inproceedings{liu2023fingpt,
  title={{FinGPT}: Democratizing internet-scale data for financial large language models},
  author={Liu, Xiao-Yang and Wang, Guoxuan and Yang, Hongyang and Zha, Daochen},
  booktitle={Proceedings of the Workshop on Instruction Tuning and Instruction Following at NeurIPS 2023},
  year={2023},
  url={https://openreview.net/forum?id=5BqWC1Fz8F}
}

@inproceedings{min2023factscore,
  title={FActScore: Fine-grained Atomic Evaluation of Factual Precision in Long Form Text Generation},
  author={Min, Sewon and Krishna, Kalpesh and Lyu, Xinxi and Lewis, Mike and Yih, Wen-tau and Koh, Pang Wei and Iyyer, Mohit and Zettlemoyer, Luke and Hajishirzi, Hannaneh},
  booktitle={Proceedings of the 2023 Conference on Empirical Methods in Natural Language Processing},
  pages={12076--12100},
  year={2023}
}

@article{petty2025codepretrain,
  title={How Does Code Pretraining Affect Language Model Task Performance?},
  author={Petty, Jackson and van Steenkiste, Sjoerd and Linzen, Tal},
  journal={Transactions on Machine Learning Research},
  year={2025},
  note={arXiv:2409.04556}
}

@article{wu2023bloomberggpt,
  title={{BloombergGPT}: A Large Language Model for Finance},
  author={Wu, Shijie and Irsoy, Ozan and Lu, Steven and Dabravolski, Vadim and Dredze, Mark and Gehrmann, Sebastian and Kambadur, Prabhanjan and Rosenberg, David and Mann, Gideon},
  journal={arXiv preprint arXiv:2303.17564},
  year={2023},
  doi={10.48550/arXiv.2303.17564}
}

@article{araci2019finbert,
  title={{FinBERT}: Financial Sentiment Analysis with Pre-trained Language Models},
  author={Araci, Dogu},
  journal={arXiv preprint arXiv:1908.10063},
  year={2019}
}

@inproceedings{xie2024finben,
  title={{FinBen}: A Holistic Financial Benchmark for Large Language Models},
  author={Xie, Qianqian and Han, Weiguang and Chen, Zhengyu and Xiang, Ruoyu and Zhang, Xiao and He, Yueru and Xiao, Mengxi and Li, Dong and Dai, Yongfu and Feng, Duanyu and Xu, Yijing and Kang, Haoqiang and Kuang, Ziyan and Yuan, Chenhan and Yang, Kailai and Luo, Zheheng and Zhang, Tianlin and Liu, Zhiwei and Xiong, Guojun and Deng, Zhiyang and Jiang, Yuechen and Yao, Zhiyuan and Li, Haohang and Yu, Yangyang and Hu, Gang and Huang, Jiajia and Liu, Xiao-Yang and Lopez-Lira, Alejandro and Wang, Benyou and Lai, Yanzhao and Wang, Hao and Peng, Min and Ananiadou, Sophia and Huang, Jimin},
  booktitle={Advances in Neural Information Processing Systems},
  volume={37},
  year={2024},
  doi={10.52202/079017-3033}
}

@inproceedings{peng2025multifinben,
  title={{MultiFinBen}: Benchmarking Large Language Models for Multilingual and Multimodal Financial Application},
  author={Peng, Xueqing and Qian, Lingfei and Wang, Yan and Xiang, Ruoyu and He, Yueru and Ren, Yang and Jiang, Mingyang and Zhang, Vincent Jim and Guo, Yuqing and Zhao, Jeff and He, Huan and Han, Yi and Feng, Yun and Jiang, Yuechen and Cao, Yupeng and Li, Haohang and Yu, Yangyang and Wang, Xiaoyu and Gao, Penglei and Lin, Shengyuan and Wang, Keyi and Yang, Shanshan and Zhao, Yilun and Liu, Zhiwei and Lu, Peng and Huang, Jerry and Wang, Suyuchen and Papadopoulos, Triantafillos and Giannouris, Polydoros and Soufleri, Efstathia and Chen, Nuo and Deng, Zhiyang and Fu, Heming and Zhao, Yijia and Lin, Mingquan and Qiu, Meikang and Smith, Kaleb E. and Cohan, Arman and Liu, Xiao-Yang and Huang, Jimin and Xiong, Guojun and Lopez-Lira, Alejandro and Chen, Xi and Tsujii, Junichi and Nie, Jian-Yun and Ananiadou, Sophia and Xie, Qianqian},
  booktitle={Proceedings of the 64th Annual Meeting of the Association for Computational Linguistics (Volume 1: Long Papers)},
  pages={16934--16963},
  publisher={Association for Computational Linguistics},
  address={San Diego, California, United States},
  doi={10.18653/v1/2026.acl-long.770},
  url={https://aclanthology.org/2026.acl-long.770/},
  year={2026}
}

@inproceedings{wei2024longform,
  title={Long-form factuality in large language models},
  author={Wei, Jerry and Yang, Chengrun and Song, Xinying and Lu, Yifeng and Hu, Nathan and Huang, Jie and Tran, Dustin and Peng, Daiyi and Liu, Ruibo and Huang, Da and Du, Cosmo and Le, Quoc V.},
  booktitle={Advances in Neural Information Processing Systems},
  volume={37},
  year={2024},
  url={https://arxiv.org/abs/2403.18802}
}

@inproceedings{song2024veriscore,
  title={{VeriScore}: Evaluating the factuality of verifiable claims in long-form text generation},
  author={Song, Yixiao and Kim, Yekyung and Iyyer, Mohit},
  booktitle={Findings of the Association for Computational Linguistics: EMNLP 2024},
  year={2024},
  address={Miami, Florida, USA},
  publisher={Association for Computational Linguistics},
  pages={9447--9474},
  doi={10.18653/v1/2024.findings-emnlp.552}
}

@article{hu2025finsearchcomp,
  title={{FinSearchComp}: Towards a Realistic, Expert-Level Evaluation of Financial Search and Reasoning},
  author={Hu, Liang and Jiao, Jianpeng and Liu, Jiashuo and Ren, Yanle and Wen, Zhoufutu and Zhang, Kaiyuan and Zhang, Xuanliang and Gao, Xiang and He, Tianci and Hu, Fei and Liao, Yali and Wang, Zaiyuan and Yang, Chenghao and Yang, Qianyu and Yin, Mingren and Zeng, Zhiyuan and Zhang, Ge and Zhang, Xinyi and Zhao, Xiying and Zhu, Zhenwei and Namkoong, Hongseok and Huang, Wenhao and Tang, Yuwen},
  journal={arXiv preprint arXiv:2509.13160},
  year={2025},
  url={https://arxiv.org/abs/2509.13160}
}

@misc{anthropic2026financial,
  author       = {Anthropic},
  title        = {Claude for Financial Services},
  year         = {2026},
  howpublished = {\url{https://github.com/anthropics/financial-services}},
  note         = {GitHub repository, accessed 2026-08-20}
}

@article{islam2023financebench,
  title={{FinanceBench}: A New Benchmark for Financial Question Answering},
  author={Islam, Pranab and Kannappan, Anand and Kiela, Douwe and Qian, Rebecca and Scherrer, Nino and Vidgen, Bertie},
  journal={arXiv preprint arXiv:2311.11944},
  year={2023}
}

@article{liu2024fingpthpc,
  title={{FinGPT-HPC}: Efficient Pretraining and Finetuning Large Language Models for Financial Applications with High-Performance Computing},
  author={Liu, Xiao-Yang and Zhang, Jie and Wang, Guoxuan and Tong, Weiqing and Walid, Anwar},
  journal={arXiv preprint arXiv:2402.13533},
  year={2024}
}

@article{xiao2026financeharness,
  title={{FinanceHarness}: Autonomous Financial Deep Research Framework},
  author={Xiao, Yijia and Han, Rujun and Chen, Yanfei and Wang, Zifeng and Jiang, Ke and CuiZhu, Zhongying and Tirumalashetty, Vishy and Wang, Wei and Gokturk, Burak and Pfister, Tomas and Lee, Chen-Yu},
  journal={arXiv preprint arXiv:2607.27853},
  year={2026}
}

@article{hall2026verifin,
  title={{VeriFin}: A Neurosymbolic Framework for Verifying {LLM}-Generated Financial Claims},
  author={Hall, Bethel and Shome, Sachi and Eiers, William},
  journal={arXiv preprint arXiv:2608.10213},
  year={2026}
}

@article{choi2026humanai,
  title={Human + {AI} in Accounting: Early Evidence from the Field},
  author={Choi, Jung Ho and Xie, Chloe L.},
  journal={Journal of Accounting Research},
  volume={64},
  number={3},
  pages={1333--1373},
  year={2026},
  doi={10.1111/1475-679X.70052}
}

@misc{churchman2026building,
  author={Churchman, Chris},
  title={Building {AI} Systems for Capital Markets},
  year={2026},
  howpublished={Goldman Sachs Exchanges podcast, interview by Allison Nathan and George Lee, August 24, 2026. \url{https://www.goldmansachs.com/insights/goldman-sachs-exchanges/building-ai-systems-for-capital-markets}},
  note={Accessed: 2026-10-12}
}

@misc{liu2026nocertificate,
  title         = {No Certificate, No Execution: Certified Traces as a Foundation for Trustworthy AI Agents},
  author        = {Liu, Xiao-Yang and Wang, Xiaodong and Capponi, Agostino},
  year          = {2026},
  eprint        = {2605.24462},
  archivePrefix = {arXiv},
  primaryClass  = {cs.AI},
  url           = {https://arxiv.org/abs/2605.24462}
}

\clearpage 

\appendix

\section{Benchmark Company Roster}
\label{app:company_roster}

Table~\ref{tab:company_roster} lists the complete 30-company benchmark roster described in Section~\ref{sec:company_cases}.

\begin{table}[htbp]
\centering
\small
\caption{Benchmark company roster.}
\label{tab:company_roster}
\begin{tabular}{lll}
\toprule
Category & Company & Ticker \\
\midrule
Information Technology & Apple & AAPL \\
Information Technology & Microsoft & MSFT \\
Information Technology & Oracle & ORCL \\
Semiconductors \& AI Infrastructure & NVIDIA & NVDA \\
Semiconductors \& AI Infrastructure & Intel & INTC \\
Semiconductors \& AI Infrastructure & Broadcom & AVGO \\
Communication Services & Alphabet & GOOGL \\
Communication Services & Meta Platforms & META \\
Communication Services & Netflix & NFLX \\
Financials & JPMorgan Chase & JPM \\
Financials & Goldman Sachs & GS \\
Financials & BlackRock & BLK \\
Health Care & Johnson \& Johnson & JNJ \\
Health Care & Eli Lilly & LLY \\
Health Care & UnitedHealth Group & UNH \\
Consumer Discretionary & Amazon & AMZN \\
Consumer Discretionary & Tesla & TSLA \\
Consumer Discretionary & Home Depot & HD \\
Consumer Staples & Walmart & WMT \\
Consumer Staples & Procter \& Gamble & PG \\
Consumer Staples & Coca-Cola & KO \\
Industrials & Caterpillar & CAT \\
Industrials & Boeing & BA \\
Industrials & Honeywell & HON \\
Energy & ExxonMobil & XOM \\
Energy & Chevron & CVX \\
Energy & ConocoPhillips & COP \\
Materials & Freeport-McMoRan & FCX \\
Materials & Dow Inc. & DOW \\
Materials & Newmont & NEM \\
\bottomrule
\end{tabular}
\end{table}

\section{Experimental Configuration and Cost Profile}
\label{app:cost_profile}
Because model capabilities and inference prices change over time, we record the model configuration and approximate token cost of the memo-generation stage. This cost profile is intended to make the benchmark easier to reproduce and to clarify the model scale represented by the reported results.

The evaluation uses four report-generation agents: Codex powered by GPT-5.5, Claude powered by Claude Sonnet 4.6, DeepSeek powered by DeepSeek-V4-Pro, and Qwen powered by qwen3.8-flash. All four agents are evaluated under the same source-bounded prompting setup. For each memo, the agent reads the memo-generation prompt together with the corresponding company-period source bundle and generates one structured analyst-style investment memo. The prompt constrains each memo to 900--1,200 words, excluding the key metrics table, and requires the same seven-section report structure across all companies, quarters, and agents.

For cost estimation, we compute token usage at the memo level and aggregate across the 360 company-period cases generated by each agent. The estimated memo-generation workload is 105.9 million input tokens and 0.46 million output tokens per agent. The input tokens are much larger than the output tokens because each memo is generated from a company-period source bundle rather than from a short prompt alone. Table~\ref{tab:cost_profile} reports the resulting API-equivalent cost estimate using model prices available at the time of the experiment.

\begin{table}[H]
\centering
\footnotesize
\caption{Estimated token usage and API-equivalent cost for memo generation.}
\label{tab:cost_profile}
\begin{tabular}{lrrrrr}
\toprule
\textbf{Agent} &
\textbf{Model} &
\textbf{Input Tokens} &
\textbf{Output Tokens} &
\textbf{Pricing Basis} &
\textbf{Est. Cost} \\
\midrule
Codex
& GPT-5.5
& 105.9M
& 0.46M
& \$5/MTok input, \$30/MTok output
& \$543 \\

Claude
& Sonnet 4.6
& 105.9M
& 0.46M
& \$3/MTok input, \$15/MTok output
& \$325 \\

DeepSeek
& DeepSeek-V4-Pro
& 105.9M
& 0.46M
& \$0.7/MTok input, \$2/MTok output
& \$71 \\

Qwen
& qwen3.8-flash
& 105.9M
& 0.46M
& \$0.15/MTok input, \$0.47/MTok output
& \$16 \\

\bottomrule
\end{tabular}
\end{table}

Under this estimate, generating 360 memos costs approximately \$543 with GPT-5.5, \$325 with Claude Sonnet 4.6, \$71 with DeepSeek-V4-Pro, and \$16 with qwen3.8-flash. Across the four-agent evaluation, the memo-generation stage uses approximately 423.6 million input tokens and 1.84 million output tokens, corresponding to an estimated combined generation cost of approximately \$955. These values should be interpreted as an implementation-specific cost profile rather than a fixed cost of the proposed validation framework, since model prices, context-window pricing, and inference settings may change over time.

\section{Claim Density and Evaluation Design}
\label{app:claim_density}

The common prompt specifies a 900--1,200-word target and analyst-style structure, but it does not ensure identical actual lengths or force agents to express the same number of claims. Codex produces 27.0\% more extracted claims per memo than Claude, 30.7\% more than DeepSeek, and 18.7\% more than Qwen. Qwen averages 53.3 claims per memo, compared with 63.2 for Codex, 49.8 for Claude, and 48.4 for DeepSeek. These comparisons describe claims per memo, not claims per word, and therefore do not isolate the effect of actual memo length. That difference is itself relevant to deployment: greater claim density creates more factual commitments for a reader to rely on and for a validation layer to check.

\section{Generation Prompt}
\label{app:prompt}

The following prompt template was used for investment memo generation.

\begin{tcolorbox}[
enhanced,
breakable,
colback=blue!3,
colframe=blue!50!black,
boxrule=0.5pt,
arc=1mm,
left=3mm,
right=3mm,
top=2mm,
bottom=2mm]

\textbf{\large Prompt for Generating Investment Memos}

\vspace{0.8em}

\textbf{Task}

You are an investment research memo generation agent for a financial LLM validation benchmark. Generate one source-grounded investment research memo for a single company-period using only the provided official source bundle.

\vspace{0.8em}

\textbf{Source Boundary}

Use only the files contained in the target company-period source bundle. Do not use information from other companies or reporting periods, outside knowledge, web search, analyst reports, news articles, market prices, or model memory. Every material claim must be supported by the provided source files.

\vspace{0.8em}

\textbf{Source Priority}

When multiple official documents are available, prioritize evidence in the following order:

\begin{enumerate}
\item SEC Form 10-Q or 10-K filings.
\item Earnings releases.
\item Financial supplements or earnings presentations.
\item Earnings call transcripts.
\end{enumerate}

If conflicting figures are encountered, prioritize SEC filings and official earnings releases. If information is unavailable, explicitly state that it is not disclosed rather than inferring unsupported facts.

\vspace{0.8em}

\textbf{Memo Requirements}

Generate a concise analyst-style investment memo. Every material numerical claim must be traceable to the source bundle. Clearly distinguish reported facts from analytical interpretation, identify GAAP and non-GAAP metrics where appropriate, and use causal language only when explicitly supported by the source materials. Do not generate unsupported financial figures, investment ratings, valuation estimates, or price targets.

\vspace{0.8em}

\textbf{Output Structure}

The generated memo should contain the following sections:

\begin{enumerate}

\item \textbf{Executive Summary / Investment View}

Summarize the overall investment view for the reporting period. Highlight the most important financial developments, major performance drivers, and primary watchpoints without providing investment ratings or price targets.

\item \textbf{Company and Segment Context}

Briefly introduce the company's principal businesses and relevant operating segments. State the fiscal reporting period and provide only source-grounded background necessary to interpret the reported results.

\item \textbf{Latest Quarter Performance}

Summarize the reported quarterly financial performance, including key metrics such as revenue, profitability, earnings, cash flow, and major segment performance where available. Report comparisons only when directly supported by the source materials.

\item \textbf{Key Metrics Table}

Present a concise table summarizing the most important reported financial metrics, including current-period values, comparison-period values, reported changes, and corresponding source references where appropriate.

\item \textbf{Drivers and Interpretation}

Explain the principal business drivers reflected in the reported results while clearly distinguishing factual evidence from analytical interpretation. Any causal explanation should be explicitly supported by the provided source materials.

\item \textbf{Risks and Watchpoints}

Identify several source-grounded risks or operational issues that should be monitored by analysts or investors. Avoid introducing speculative or unsupported risk factors.

\item \textbf{Source-Grounded Conclusion}

Provide a concise overall assessment of the reporting period using one of three evidence-based tones (\emph{constructive}, \emph{mixed}, or \emph{cautious}). Do not provide valuation, investment recommendations, or price targets.

\end{enumerate}

\end{tcolorbox}

\clearpage
\subsection{Relationship to Production Financial Agent Frameworks}
The memo-generation task studied in this paper is not just an academic exercise: it closely mirrors workflows already shipped in production-oriented agent frameworks. Anthropic's own reference implementation for financial services~\citep{anthropic2026financial} includes an \textit{Earnings Reviewer} agent whose stated workflow is ``earnings call and filings $\rightarrow$ model update $\rightarrow$ note draft,'' which is functionally the same source-to-memo pipeline evaluated here. This is a curated collection of reference agents, skills, and data connectors rather than a benchmark, so it offers no accuracy metrics or evaluation set to compare against directly; we can only compare the two approaches structurally, not with numbers.

The repository is explicit that its agents draft analyst work product for human review rather than final output: every generated artifact is described as ``staged for human sign-off,'' and the repository's own documentation states that its agents ``do not make investment recommendations'' on their own authority. Beyond generic presentation and spreadsheet quality-control skills, the framework does not include a systematic, source-grounded mechanism for checking whether the individual factual claims in a generated note are actually supported by the underlying earnings materials; reliability instead depends entirely on manual analyst review of each output. This matches the practical concern motivating our work: production systems built for exactly this task currently treat full manual re-verification as the only safeguard against factual errors. The claim-level extraction and verification framework proposed in this paper is intended as a complementary layer for this setting, one that shows which specific claims in a generated memo are source-supported, contradicted, or unverifiable before a human reviewer sees the report, rather than requiring that reviewer to re-check every claim from scratch.

\section{Example Generated Investment Memo}
\label{app:memo_example}

The following example presents a representative investment research memo generated for The Home Depot, Inc. (HD) FY2024 Q4 using GPT-5-Codex.

\begin{tcolorbox}[
enhanced,
breakable,
colback=blue!3,
colframe=blue!50!black,
boxrule=0.5pt,
arc=1mm,
left=3mm,
right=3mm,
top=2mm,
bottom=2mm]

\small

{\Large\textbf{The Home Depot, Inc. (HD) Investment Memo -- FY2024 Q4}}

\bigskip

\textbf{1. Executive Summary / Investment View}

The source-grounded investment view for Home Depot in FY2024 Q4 is \textbf{constructive}. Home Depot reported net sales of \$39.7 billion, up 14.1\% year over year, with gross profit of \$13.0 billion, operating income of \$4.5 billion, net earnings of \$3.0 billion, and diluted EPS of \$3.02. Customer transactions were 400.4 million, average ticket was \$89.11, and sales per retail square foot were \$556.90. The main watchpoints are comparable sales, customer transactions, average ticket, gross margin, SG\&A leverage, Pro and DIY demand mix, acquired-business contribution, inventory, cash flow, capital allocation, and housing or repair-and-remodel demand. This memo uses only files in the target quarter folder and provides no rating or price target.

\medskip

\textbf{2. Company and Segment Context}

Home Depot is a home improvement retailer with sales across building materials, decor, hardlines, and services, spanning geographic markets in and outside the United States. FY2024 Q4 is mapped by the source README to calendar 2025 Q1 and ended on February 2, 2025. The source files disclose consolidated sales, gross profit, operating income, net earnings, EPS, selected sales data, cash flows, and detailed sales lines where applicable, and this memo follows the labels used in the target-quarter source files.

\medskip

\textbf{3. Latest Quarter Performance}

Net sales were \$39.7 billion, compared with \$34.8 billion in the prior-year period. Gross profit was \$13.0 billion, compared with \$11.5 billion; operating income was \$4.5 billion, compared with \$4.1 billion; net earnings were \$3.0 billion, compared with \$2.8 billion; and diluted EPS was \$3.02, compared with \$2.82. Customer transactions were 400.4 million, average ticket was \$89.11, and sales per retail square foot were \$556.90. Building Materials sales were \$52.8 billion, up 0.2\% year over year; Decor sales were \$51.8 billion, down 0.3\%; and Hardlines sales were \$48.6 billion, up 1.0\%. The largest disclosed positive swing was net sales, and no listed driver declined year over year.

\medskip

\textbf{4. Key Metrics Table}

\begin{center}
\footnotesize
\setlength{\tabcolsep}{6pt}
\begin{tabular}{@{}llll@{}}
\toprule
Metric & Current & Prior Year & Change\\
\midrule
Net Sales & \$39.7B & \$34.8B & +14.1\%\\
Gross Profit & \$13.0B & \$11.5B & +13.3\%\\
Gross Margin & 32.8\% & 33.1\% & Not disclosed\\
Operating Income & \$4.5B & \$4.1B & +8.5\%\\
Net Earnings & \$3.0B & \$2.8B & +7.0\%\\
Diluted EPS & \$3.02 & \$2.82 & +7.1\%\\
Comparable Sales Growth & Not disclosed & Not disclosed & Not disclosed\\
Customer Transactions & 400.4M & 372.0M & +7.6\%\\
Average Ticket & \$89.11 & \$88.87 & +0.3\%\\
Sales per Retail Sq. Ft. & \$556.90 & \$550.50 & +1.2\%\\
U.S. Sales & Not disclosed & Not disclosed & Not disclosed\\
Outside-U.S. Sales & Not disclosed & Not disclosed & Not disclosed\\
Operating Cash Flow & \$19.8B & \$21.2B & --6.4\%\\
\bottomrule
\end{tabular}
\end{center}

\medskip

\textbf{5. Drivers and Interpretation}

The clearest source-grounded driver evidence comes from the income statement, selected sales data, cash flow table, and detailed sales tables. Net sales changed from \$34.8 billion to \$39.7 billion, customer transactions were 400.4 million where disclosed, and average ticket was \$89.11; comparable sales growth itself was not disclosed in the source files. The largest disclosed positive swing was net sales, and no listed driver declined year over year. The memo avoids external assumptions about housing turnover, interest rates, consumer credit, Pro backlog, weather, lumber prices, or valuation unless the target-quarter folder directly supports them.

\medskip

\textbf{6. Risks and Watchpoints}

Several operational risks remain relevant. First, comparable sales and transaction trends matter because they separate organic retail momentum from calendar, acquisition, or footprint effects. Second, average ticket and product mix can offset or amplify transaction weakness. Third, gross margin and SG\&A leverage affect operating income when sales growth is modest. Fourth, cash flow and capital allocation matter because Home Depot returns capital through dividends and repurchases while also investing in stores, supply chain, technology, and acquisitions. Finally, housing, repair-and-remodel, and Pro demand cycles remain important operating risks, though this memo does not import external macro forecasts.

\medskip

\textbf{7. Source-Grounded Conclusion}

Overall, the evidence supports a \textbf{constructive} assessment for FY2024 Q4. Home Depot reported \$39.7 billion of net sales, \$13.0 billion of gross profit, \$4.5 billion of operating income, \$3.0 billion of net earnings, and diluted EPS of \$3.02, with average ticket of \$89.11; comparable sales growth was not disclosed in the source files. The operating story is the balance between sales growth, comparable sales, transactions, ticket, gross margin, operating leverage, and cash flow. Based solely on the official company-period source bundle, the provided files support an operating assessment focused on sales, earnings, selected sales data, detailed sales lines, and cash flow, but they do not support a valuation conclusion, price target, or personalized investment recommendation.

\end{tcolorbox}

\subsection{Claim Verification Pipeline}
\label{app:pipeline_figure}

\begin{figure}[h]
  \centering
  \includegraphics[width=0.75\textwidth]{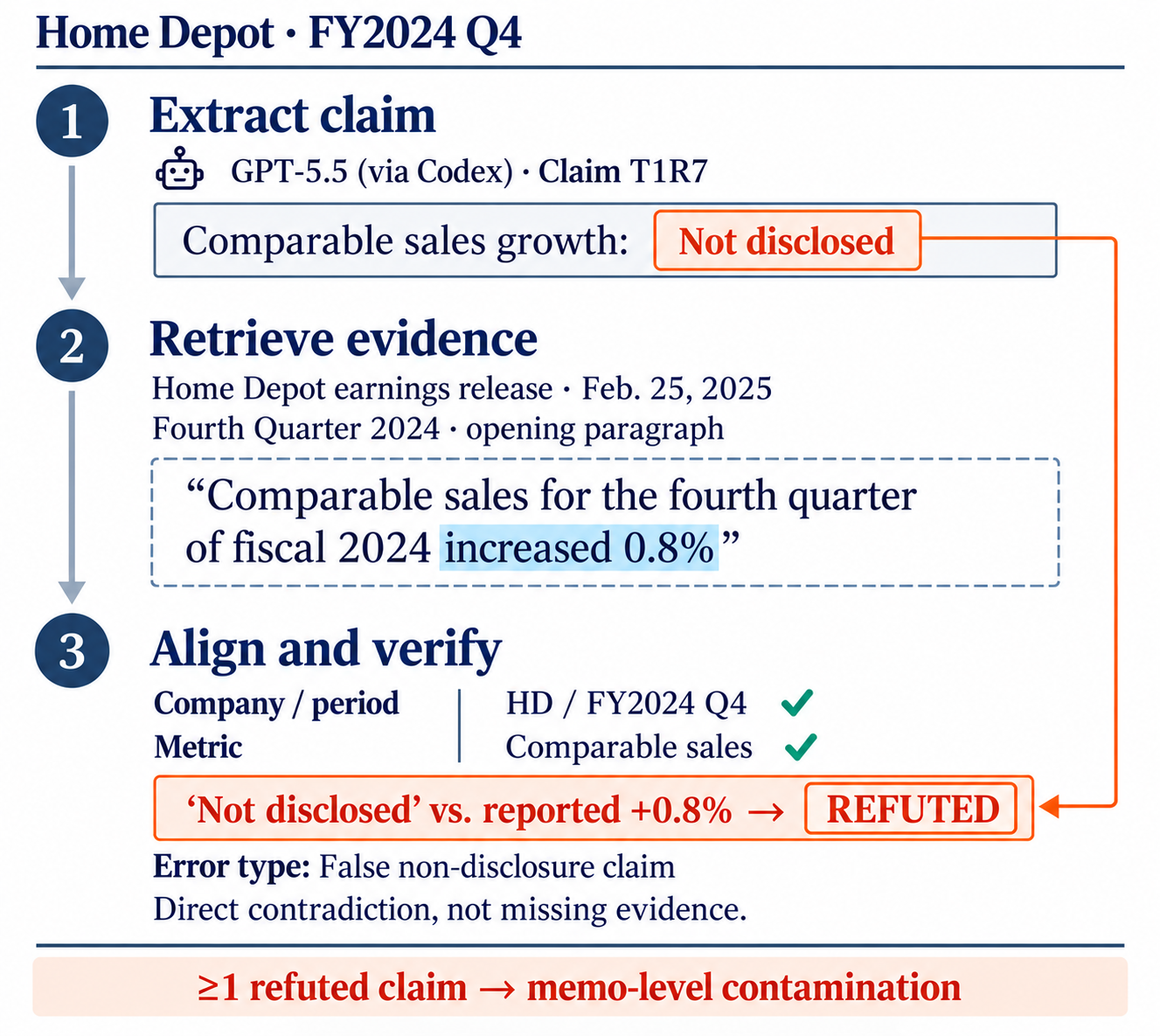}
  \caption{We evaluate LLM-generated investment research memos through three stages: claim extraction, evidence retrieval, and alignment and verification. Full memo and claim labels appear in Appendix~\ref{app:memo_example}.}
  \label{fig:hd_validation_case}
\end{figure}

Figure~\ref{fig:hd_validation_case} illustrates the claim extraction, evidence retrieval, and alignment-and-verification pipeline using the Home Depot FY2024~Q4 example discussed in the introduction (full memo and claim labels appear in Appendix~\ref{app:memo_example}).

\subsection{Annotated Claim Extraction}
\label{sec:hd_annotated_claims}
To make the connection between a generated memo and its downstream verification fully concrete, Table~\ref{tab:hd_claim_table} lists every one of the 47 claims extracted by the tagging module from the Home Depot memo shown above, ordered to match the memo's own section structure. For each claim we report the verification outcome assigned by the pipeline and a brief statement of the evidentiary basis for that outcome. Of the 47 claims, 14 are verified, 11 are partially verified, 4 are unverified, 11 are not applicable, and 7 are refuted --- all five outcomes therefore appear within this single memo. Appendix~\ref{app:tag_example} walks through one of the refuted rows, \texttt{T1R7}, in full JSON detail.

{\footnotesize
\begin{longtable}{@{}p{0.11\linewidth}p{0.44\linewidth}p{0.09\linewidth}p{0.30\linewidth}@{}}
\caption{Every claim extracted from the Home Depot FY2024 Q4 memo in Appendix~\ref{app:memo_example}, with verification outcome and basis. All claim IDs share the prefix \texttt{HD\_FY2024\_Q4}.}
\label{tab:hd_claim_table} \\
\toprule
\textbf{ID} & \textbf{Claim Text} & \textbf{Outcome} & \textbf{Basis} \\
\midrule
\endfirsthead
\toprule
\textbf{ID} & \textbf{Claim Text} & \textbf{Outcome} & \textbf{Basis} \\
\midrule
\endhead
\midrule
\multicolumn{4}{r}{\textit{continued on next page}} \\
\endfoot
\bottomrule
\endlastfoot
\multicolumn{4}{l}{\textit{1. Executive Summary / Investment View}} \\
\midrule
\texttt{P2S1C1} & The source-grounded investment view for Home Depot in FY2024 Q4 is constructive. & N/A & Not a factual claim (interpretation/disclaimer). \\
\texttt{P2S2C2} & Home Depot reported net sales were \$39.7 billion. & Verified & Value matched official source. \\
\texttt{P2S2C3} & Home Depot reported net sales increased 14.1\% year over year. & Verified & Value matched official source. \\
\texttt{P2S3C4} & Gross profit was \$13.0 billion, operating income was \$4.5 billion, net earnings were \$3.0 billion, and diluted EPS was \$3.02. Customer transactions were 400.4 million. & Partial & 3/5 figures matched; remainder unconfirmed. \\
\texttt{P2S3C5} & Average ticket was \$89.11. & Unverified & No matching evidence retrieved. \\
\texttt{P2S3C6} & Sales per retail square foot were \$556.90. & Refuted & Value does not match official source (outside tolerance). \\
\texttt{P2S4C7} & The main watchpoints are comparable sales, customer transactions, average ticket, gross margin, SG\&A leverage, Pro and DIY demand mix, acquired-business contribution, inventory, cash flow, capital allocation, and housing or repair-and-remodel demand. & N/A & Not a factual claim (interpretation/disclaimer). \\
\texttt{P2S5C8} & This memo uses only files in the target quarter folder and provides no rating or price target. & N/A & Not a factual claim (interpretation/disclaimer). \\
\addlinespace
\multicolumn{4}{l}{\textit{2. Company and Segment Context}} \\
\midrule
\texttt{P3S1C9} & Home Depot is a home improvement retailer with sales across building materials, decor, hardlines, services, and geographic markets in and outside the United States. & Partial & Related evidence found; not a confirmed match. \\
\texttt{P3S2C10} & FY2024 Q4 is mapped by the README to Cal 2025 Q1 and ended on February 2, 2025. & Unverified & No matching evidence retrieved. \\
\texttt{P3S3C11} & The source files disclose consolidated sales, gross profit, operating income, net earnings, EPS, selected sales data, cash flows, and detailed sales lines where applicable. & N/A & Not a factual claim (interpretation/disclaimer). \\
\texttt{P3S4C12} & This memo follows the labels used in the target quarter source files. & N/A & Not a factual claim (interpretation/disclaimer). \\
\addlinespace
\multicolumn{4}{l}{\textit{3. Latest Quarter Performance}} \\
\midrule
\texttt{P4S1C13} & Net sales were \$39.7 billion, compared with \$34.8 billion in the prior-year period. & Verified & All values matched official source. \\
\texttt{P4S2C14} & Gross profit was \$13.0 billion, compared with \$11.5 billion. & Verified & All values matched official source. \\
\texttt{P4S2C15} & Operating income was \$4.5 billion, compared with \$4.1 billion. & Verified & All values matched official source. \\
\texttt{P4S2C16} & Net earnings were \$3.0 billion, compared with \$2.8 billion. & Verified & Value matched official source. \\
\texttt{P4S2C17} & Diluted EPS was \$3.02, compared with \$2.82. Customer transactions were 400.4 million. & Partial & 2/3 figures matched; remainder unconfirmed. \\
\texttt{P4S2C18} & Average ticket was \$89.11. & Unverified & No matching evidence retrieved. \\
\texttt{P4S2C19} & Sales per retail square foot were \$556.90. & Refuted & Value does not match official source (outside tolerance). \\
\texttt{P4S3C20} & Building Materials sales were \$52.8 billion, up 0.2\% year over year. & Partial & 1/2 figures matched; remainder unconfirmed. \\
\texttt{P4S3C21} & Decor sales were \$51.8 billion, down 0.3\% year over year. & Partial & 1/2 figures matched; remainder unconfirmed. \\
\texttt{P4S3C22} & Hardlines sales were \$48.6 billion, up 1.0\% year over year. & Partial & 1/2 figures matched; remainder unconfirmed. \\
\texttt{P4S4C23} & The largest disclosed positive swing was net sales; no listed driver declined year over year. & Partial & Related evidence found; not a confirmed match. \\
\addlinespace
\multicolumn{4}{l}{\textit{4. Key Metrics Table}} \\
\midrule
\texttt{T1R1} & Net sales were \$39.7 billion in Current Period, with change of up 14.1\%. & Verified & All values matched official source. \\
\texttt{T1R2} & Gross profit was \$13.0 billion in Current Period, with change of up 13.3\%. & Verified & All values matched official source. \\
\texttt{T1R3} & Gross margin was 32.8\% in Current Period. & Verified & All values matched official source. \\
\texttt{T1R4} & Operating income was \$4.5 billion in Current Period, with change of up 8.5\%. & Verified & All values matched official source. \\
\texttt{T1R5} & Net earnings was \$3.0 billion in Current Period, with change of up 7.0\%. & Verified & Value matched official source. \\
\texttt{T1R6} & Diluted EPS was \$3.02 in Current Period, with change of up 7.1\%. & Verified & All values matched official source. \\
\texttt{T1R7} & Comparable sales growth was not disclosed in Current Period. & Refuted & Claim asserts non-disclosure; official source discloses this metric. \\
\texttt{T1R8} & Customer transactions was 400.4 million in Current Period, with change of up 7.6\%. & Partial & Numeric mismatch, but evidence too ambiguous for a confident refutation. \\
\texttt{T1R9} & Average ticket was \$89.11 in Current Period, with change of up 0.3\%. & Unverified & No matching evidence retrieved. \\
\texttt{T1R10} & Sales per retail square foot was \$556.90 in Current Period, with change of up 1.2\%. & Partial & 1/2 figures matched; remainder unconfirmed. \\
\texttt{T1R11} & U.S. sales were not disclosed in Current Period. & Refuted & Claim asserts non-disclosure; official source discloses this metric. \\
\texttt{T1R12} & Outside-U.S. sales were not disclosed in Current Period. & Refuted & Claim asserts non-disclosure; official source discloses this metric. \\
\texttt{T1R13} & Operating cash flow was \$19.8 billion in Current Period, with change of down 6.4\%. & Verified & All values matched official source. \\
\addlinespace
\multicolumn{4}{l}{\textit{5. Drivers and Interpretation}} \\
\midrule
\texttt{P5S2C24} & Net sales changed from \$34.8 billion to \$39.7 billion. & Verified & All values matched official source. \\
\texttt{P5S3C25} & Comparable sales changed not disclosed, customer transactions were 400.4 million where disclosed, and average ticket was \$89.11. & Refuted & Claim asserts non-disclosure; official source discloses this metric. \\
\texttt{P5S4C26} & The largest disclosed positive swing was net sales; no listed driver declined year over year. & Partial & Related evidence found; not a confirmed match. \\
\texttt{P5S5C27} & The memo avoids external assumptions about housing turnover, interest rates, consumer credit, Pro backlog, weather, lumber prices, or valuation unless the target quarter folder directly supports them. & N/A & Not a factual claim (interpretation/disclaimer). \\
\addlinespace
\multicolumn{4}{l}{\textit{6. Risks and Watchpoints}} \\
\midrule
\texttt{P6S3C28} & Gross margin and SG\&A leverage affect operating income when sales growth is modest. & Partial & Related evidence found; not a confirmed match. \\
\texttt{P6S5C29} & Housing, repair-and-remodel, and Pro demand cycles remain important operating risks, but this memo does not import external macro forecasts. & N/A & Not a factual claim (interpretation/disclaimer). \\
\addlinespace
\multicolumn{4}{l}{\textit{7. Source-Grounded Conclusion}} \\
\midrule
\texttt{P7S1C30} & The conclusion for FY2024 Q4 is constructive. & N/A & Not a factual claim (interpretation/disclaimer). \\
\texttt{P7S2C31} & Home Depot reported \$39.7 billion of net sales, \$13.0 billion of gross profit, \$4.5 billion of operating income, \$3.0 billion of net earnings, diluted EPS of \$3.02, comparable sales growth of not disclosed, and average ticket of \$89.11. & Refuted & Claim asserts non-disclosure; official source discloses this metric. \\
\texttt{P7S3C32} & The operating story is the balance between sales growth, comparable sales, transactions, ticket, gross margin, operating leverage, and cash flow. & N/A & Not a factual claim (interpretation/disclaimer). \\
\texttt{P7S4C33} & The provided files support an operating assessment focused on sales, earnings, selected sales data, detailed sales lines, and cash flow. & N/A & Not a factual claim (interpretation/disclaimer). \\
\texttt{P7S5C34} & They do not support a valuation conclusion, price target, or personalized investment recommendation. & N/A & Not a factual claim (interpretation/disclaimer). \\
\end{longtable}
}

\subsection{Claim-Level Case Studies}
\label{sec:case_studies}
Whereas Appendix~\ref{sec:hd_annotated_claims} adjudicates every claim in a
single generated memo, including one example of each verification outcome,
this section instead illustrates each outcome in Table 6 with one
representative claim.

A verified claim requires an exact match against the source: T1R4 ("Operating
income was \$4.5 billion in Current Period, with change of up 8.5\%") matched
the corresponding 10-K figures with no ambiguity.

A partially verified claim arises when the evidence is suggestive but not
conclusive: T1R8 reports customer transactions of 400.4 million, up 7.6\%; the
retrieved figure did not exactly match, but the mismatch was judged too
ambiguous to support a confident refutation, reflecting the deliberately
conservative labeling policy.

An unverified claim occurs when no matching evidence can be located at all:
T1R9 ("Average ticket was \$89.11 in Current Period") could not be checked
because the source bundle contains no directly matching figure for this
metric.

A not-applicable claim has no single factual value to check against the
source: P2S1C1 ("The source-grounded investment view for Home Depot in FY2024
Q4 is constructive.") is an interpretive judgment rather than a checkable
statement, and is excluded from the error-rate calculation for that reason.

A refuted claim requires direct contradiction by the source: T1R7 ("Comparable
sales growth was not disclosed in Current Period") is refuted because the
source filing does in fact disclose this metric, making the claim's central
assertion false rather than merely unconfirmed.

\section{Claim Taxonomy and Tagging}
\label{app:claim_taxonomy}

\subsection{Claim and Verification Types}
\label{app:claim_types}
The extraction framework assigns two related labels to each extracted claim. The first is the \textbf{claim kind}, which describes the semantic form of the memo statement. The second is the \textbf{verification type}, which specifies the evidence operation required for downstream validation.

Claim kinds are defined as follows:
\begin{itemize}
\item \textbf{Quantitative claim}: reports a financial or operating value for the target period.
\item \textbf{Comparative numeric claim}: states a numerical change across periods, such as year-over-year growth or sequential movement.
\item \textbf{Non-numeric comparative claim}: compares business performance, segment trends, or operating conditions without stating a specific computed value.
\item \textbf{Causal claim}: attributes performance to specific drivers using language such as ``driven by,'' ``due to,'' or ``supported by.''
\item \textbf{Management guidance or outlook claim}: reports management forecasts, targets, guidance, or forward-looking commentary.
\item \textbf{Descriptive claim}: captures source-linked factual narrative about the company, segment, or reporting period.
\item \textbf{Analytical judgment}: expresses interpretation, implications, risks, or watchpoints based on the reported evidence.
\item \textbf{Source-boundary or disclaimer claim}: describes source limitations, ratings, price targets, valuation assumptions, or external-information constraints.
\end{itemize}

Verification types define the downstream evidence burden for each claim:
\begin{itemize}
\item \textbf{Current-period numeric lookup}: checks a current-period numeric value against the source bundle.
\item \textbf{Comparison checking}: verifies whether a stated comparison is supported by reported comparison-period values.
\item \textbf{Computed change}: recalculates a numerical change from values available in the source materials.
\item \textbf{Narrative attribution}: verifies whether a causal explanation is explicitly supported by the sources.
\item \textbf{Management guidance or outlook}: checks guidance, targets, or management outlook against official disclosures.
\item \textbf{Descriptive fact}: verifies factual narrative against relevant source text.
\item \textbf{Analytical judgment}: evaluates whether an interpretation is grounded in disclosed evidence.
\item \textbf{Source-boundary or disclaimer}: identifies claims about source scope, unsupported assumptions, ratings, or recommendation boundaries.
\end{itemize}

\subsection{Claim Tagging}
Each generated memo is first processed by a claim tagger that converts the Markdown report into a structured set of verification targets. The tagger extracts material claims from both prose paragraphs and key metrics tables, and emits one JSON record for each extracted claim. Each record contains the claim text, a supporting memo excerpt, and memo-location metadata such as section name, line span, paragraph position, table index, or table-row position when applicable.

Using the taxonomy defined in Appendix~\ref{app:claim_types}, the tagger assigns each claim both a claim kind and a verification type. These labels guide the initial verification route for each record. For example, a current-period revenue claim is routed to current-period numeric lookup, a period-over-period growth claim is routed to comparison checking or computed-change recalculation, and a causal statement is routed to narrative attribution checking.

\begin{table}[htbp]
\centering
\small
\caption{Example of a structured claim-tag record produced by the claim tagger.}
\label{tab:claim_tag_example_appendix}
\begin{tabularx}{\textwidth}{p{3.4cm} X}
\toprule
\textbf{Field} & \textbf{Example} \\
\midrule
Report ID & \texttt{AAPL\_FY2025\_Q4} \\
Company / Period & Apple Inc., FY2025 Q4 \\
Section & Key Metrics Table \\
Source block type & Table row \\
Claim text & Net sales were \$102.5 billion in the current period, compared with \$94.9 billion in the prior-year period, with change of up 7.9\%. \\
Memo excerpt & Metric: Net sales; Current Period: \$102.5 billion; Prior-Year / Comparison Period: \$94.9 billion; Change: up 7.9\%. \\
Claim kind & Comparative numeric claim \\
Verification type & Computed change \\
Matching strategy & Same-table recomputation \\
Structured metric & Net sales \\
Current value & \$102.5 billion \\
Prior value & \$94.9 billion \\
Reported change & +7.9\% \\
Initial verification route & Recompute the reported change from the current-period and prior-year values, then compare the underlying values and change calculation against the corresponding official source bundle. \\
\bottomrule
\end{tabularx}
\end{table}

The claim-tag JSON also records information needed for downstream verification and audit-ability, including the matching strategy, verifiability level, company and fiscal-period metadata, source files used during memo generation, extracted numeric metrics, time references, and financial entities. When available, the record also stores a structured representation of the metric, current value, prior value, and reported growth or change value. Table~\ref{tab:claim_tag_example_appendix} illustrates how a table-row statement from a generated memo is converted into a structured verification target.

In the latest four-agent evaluation, the claim-tagging stage processes 1,440 quarterly investment memos in total, including 360 Codex-generated memos, 360 Claude-generated memos, 360 DeepSeek-generated memos, and 360 Qwen-generated memos across the same 30-company sample. The tagger extracts 77,272 claims overall: 22,757 claims from Codex-generated memos, 17,920 claims from Claude-generated memos, 17,416 claims from DeepSeek-generated memos, and 19,179 claims from Qwen-generated memos. Since the four agents produce different numbers of extracted claims per memo, the downstream evaluation reports both raw counts and rate-based metrics.

\clearpage

\section{Example Claim Tag}
\label{app:tag_example}

The following example illustrates the structured claim record produced by the claim extraction and tagging module.
This record corresponds to claim \texttt{T1R7} in Table~\ref{tab:hd_claim_table} (Appendix~\ref{sec:hd_annotated_claims}); its verification output is shown next in Appendix~\ref{app:verification_example}.

\begin{tcolorbox}[
enhanced,
breakable,
colback=blue!3,
colframe=blue!50!black,
boxrule=0.5pt,
arc=1mm,
left=3mm,
right=3mm,
top=2mm,
bottom=2mm]

\small
\ttfamily

\begin{lstlisting}[basicstyle=\ttfamily\fontsize{10}{11.5}\selectfont,breaklines=true,breakatwhitespace=false,columns=fullflexible,keepspaces=true,aboveskip=0pt,belowskip=0pt]
{
  "claim_id": "HD_FY2024_Q4:T1R7",
  "report_id": "HD_FY2024_Q4",
  "company_name": "The Home Depot, Inc.",
  "ticker": "HD",
  "fiscal_period": "FY2024 Q4",
  "period_end_date": "2025-02-02",
  "calendar_period": "Cal 2025 Q1",
  "source_files_used": [
    "README.md",
    "HomeDepot_Press_Release_FY2024_Q4.htm",
    "10K-FY2024-as-filed.htm"
  ],
  "input_source_folder": "source_bundle/HomeDepot/HD_FY2024_Q4/",
  "section": "Key Metrics Table",
  "source_block_type": "table_row",
  "claim_text": "Comparable sales growth was not disclosed in Current Period.",
  "evidence_quote": "Metric: Comparable sales growth | Current Period: not disclosed | Prior-Year Period: not disclosed | Change: not disclosed | Source Note: Selected sales data",
  "claim_kind": "descriptive",
  "verification_type": "descriptive_fact",
  "matching_strategy": "source_text_lookup",
  "verifiability": "verifiable",
  "verification_status": "pending",
  "structured_claim": null,
  "metrics": [],
  "time_refs": [
    {
      "raw": "FY2024 Q4",
      "source": "generation_log"
    }
  ],
  "financial_entities": {
    "company": "The Home Depot, Inc.",
    "ticker": "HD",
    "segments_or_lines": ["Comparable", "Current Period. Comparable"],
    "metric_names": ["sales"]
  }
}
\end{lstlisting}

\end{tcolorbox}
\clearpage

\section{Example Verification Output}
\label{app:verification_example}

The following example illustrates a structured verification record for the same claim shown in Appendix~\ref{app:tag_example} and listed as \texttt{T1R7} in Table~\ref{tab:hd_claim_table} (Appendix~\ref{sec:hd_annotated_claims}), a claim that was ultimately classified as \texttt{refuted}.

\begin{tcolorbox}[
enhanced,
breakable,
colback=blue!3,
colframe=blue!50!black,
boxrule=0.5pt,
arc=1mm,
left=3mm,
right=3mm,
top=2mm,
bottom=2mm]

\small
\ttfamily

\begin{lstlisting}[basicstyle=\ttfamily\fontsize{10}{11.5}\selectfont,breaklines=true,breakatwhitespace=false,columns=fullflexible,keepspaces=true,aboveskip=0pt,belowskip=0pt]
{
  "claim_id": "HD_FY2024_Q4:T1R7",
  "original_claim_text": "Comparable sales growth was not disclosed in Current Period.",
  "status": "refuted",
  "verification_outcome_detail": "disclosure_found",
  "confidence": "high",
  "verification_type_used": "descriptive_fact",
  "matching_strategy_used": "official_fact_store",
  "tolerance_policy_applied": null,
  "recalculation": null,
  "evidence": [
    {
      "source_id": "src_hd_pres_01",
      "source_type": "press_release",
      "source_title": "HomeDepot_Press_Release_FY2024_Q4",
      "source_text": "Sales for the fourth quarter of fiscal 2024 were $39.7 billion, an increase of $4.9 billion, or 14.1% from the fourth quarter of fiscal 2023. Comparable sales for the fourth quarter of fiscal 2024 increased 0.8%, and comparable sales in the U.S. increased 1.3%.",
      "match_type": "official_fact_store_disclosure_found",
      "match_details": {
        "claim_asserts_not_disclosed": true,
        "metric_found": "net_sales",
        "source_values": [39704.0, 34786.0, 14.1, 159514.0]
      },
      "type": "contradicting"
    }
  ],
  "notes": "Claim says the metric was not disclosed, but an official source contains a matching disclosure.",
  "verified_at": "2026-08-29T01:16:24Z"
}
\end{lstlisting}

\end{tcolorbox}
\clearpage

\section{Relation to Prior Theoretical Work}
\label{app:related-theory}

A framework for certified execution of AI agents distinguishes two separate questions: whether a single action is individually permissible, and whether the complete execution trace is permissible under a governing policy system \citep{liu2026nocertificate}. Within this framework, every action in an agent's trace must carry a verifiable certificate before execution, and it is shown that a sequence of individually permissible actions does not guarantee that the overall trace is itself permissible: even when each step is certified on its own, the trace as a whole can still violate the governing policy.

Our empirical results share a related motivation but answer a different question. Section~\ref{sec:claim-memo} measures how claim-level contradictions accumulate within a complete memo: the claim-level refutation rate never exceeds 2\% across the four evaluated agents, yet 36.94--41.11\% of memos contain at least one refuted claim. Under an independence assumption, a memo with $n$ claims and per-claim error rate $p$ has contamination probability $1-(1-p)^n$, which grows quickly with $n$ even for small $p$ (Section~\ref{sec:claim-memo}). This is not a direct empirical instance of the certified-trace result above: that framework establishes a permissibility relationship between individual actions and a complete trace under a defined policy system, whereas we measure the accumulation of factual contradictions across claims within a memo, with no certification mechanism or policy system of our own. The two results are related in motivation --- both caution against judging a composite output by its parts alone --- but they are distinct contributions. A more precise connection is that our claim extraction, evidence alignment, and verification labels could supply the evaluation material for a future workflow that performs independent verification and certification before a memo is released, in the spirit of the certified-trace framework.

\section{GAAP vs. Non-GAAP Measures in SEC Filings}
\label{app:gaap-xbrl}

GAAP is an accounting basis; XBRL is a machine-readable representation of disclosures. U.S. GAAP financial statements may use standard \texttt{us-gaap} taxonomy elements and company-specific extensions where an appropriate standard element is unavailable. A missing standard tag therefore does not identify a non-GAAP measure, and a matching tag alone does not establish matching periods, units, reporting scope, or metric definitions.

Non-GAAP measures must be interpreted using the company's disclosed definition and, where available in the source bundle, reconciliation to the most directly comparable GAAP measure. Similar names need not imply identical adjustments across companies or periods. A non-GAAP value cannot be substituted directly for a GAAP value, or vice versa; any comparison must account for the disclosed adjustments. Our verifier uses source tables and text, including facts extracted into an internal store, as evidence. Accounting basis is checked separately from metric and period alignment; it is not inferred from the presence or absence of a standard XBRL tag. These checks do not amount to an audit of the issuer's accounting or a comprehensive validation of non-GAAP reconciliations.

\section{Proposed Verify-and-Revise Workflow}
\label{app:verify_revise_loop}

This appendix sketches, as a research direction, a closed loop that extends our source-bounded verification pipeline to run during memo drafting rather than only after a memo is complete. We do not implement or evaluate this loop in the current study; the sketch below is intended to make the proposal concrete enough to critique and build on.

\textbf{Proposed loop.}
\begin{enumerate}
\item \textbf{Draft claims.} The generating agent produces a memo (or a partial draft) and the claim tagger extracts its material claims, as in Section~\ref{sec:claim-memo}.
\item \textbf{Verify independently.} Each claim is checked against the source bundle using the same alignment and labeling procedure as Section~\ref{sec:source_based_verification}, yielding one of the five outcome labels.
\item \textbf{Revise unresolved or refuted claims.} Claims labeled \texttt{unverified} or \texttt{refuted} are sent back to the generating agent together with the evidence the verifier retrieved, and the agent attempts a revision grounded in that evidence.
\item \textbf{Re-verify.} Revised claims are checked again under the same procedure.
\item \textbf{Release or escalate.} A memo is released once no claim remains \texttt{refuted} (or, depending on the deployment's risk tolerance, once remaining issues fall below some materiality threshold); otherwise, the memo is escalated for human review rather than released automatically.
\end{enumerate}

\textbf{Open questions.} Realizing this loop raises design decisions this paper does not resolve:
\begin{itemize}
\item \textit{Stopping condition.} The loop cannot repeat indefinitely. A cap on the number of revision rounds per claim (or per memo) is needed, after which remaining issues are escalated rather than re-attempted.
\item \textit{Revision budget.} Each round of re-generation and re-retrieval has a cost; a budget is needed to bound how much verification and revision effort a single memo may consume before escalation becomes the cheaper option.
\item \textit{Insufficient evidence.} Some claims will remain unresolved even after revision because the source bundle simply does not contain the needed evidence, not because the agent's drafting was wrong. The loop needs an explicit fallback for this case --- for example, retaining the \texttt{unverified} label with a flag noting that revision was attempted, rather than forcing a \texttt{verified} or \texttt{refuted} label or silently dropping the claim.
\end{itemize}

Whether this loop reduces memo-level contamination in practice, and at what computational cost, is an open empirical question that our current evaluation --- which verifies memos only after they are fully written --- does not address.

\end{document}